\documentclass[preprint,12pt]{elsarticle}

\usepackage{amssymb}
\usepackage{graphicx}
\usepackage{booktabs}
\usepackage{tabularx}
\usepackage{array}
\usepackage{lscape}
\usepackage{longtable}
\usepackage{multirow}
\usepackage{amsmath}
\usepackage{colortbl}
\usepackage{subfigure}
\usepackage{url}

\newcommand{\PreserveBackslash}[1]{\let\temp=\\#1\let\\=\temp}
\newcolumntype{C}[1]{>{\PreserveBackslash\centering}p{#1}}
\newcolumntype{R}[1]{>{\PreserveBackslash\raggedleft}p{#1}}
\newcolumntype{L}[1]{>{\PreserveBackslash\raggedright}p{#1}}
\newtheorem{definition}{Definition}[section]

\journal{} 
\begin{document}

\begin{frontmatter}



\title{Multi-fractal analysis of weighted networks}


\author[address1,address2]{Daijun Wei}
\author[address3]{Xiaowu Chen}
\author[address1]{Cai Gao}
\author[address1]{Haixin Zhang}
\author[address1]{Bo Wei}
\author[address1,address4]{Yong Deng \corref{label1}}
\address[address1]{School of Computer and Information Science, Southwest University, Chongqing 400715, China}
\address[address2]{School of Science, Hubei University for Nationalities, Enshi 445000, China}
\address[address3]{School of Computer Science, BeiHang University, Beijing 100191, China}
\address[address4]{School of Engineering, Vanderbilt University, TN 37235, USA}
\cortext[label1]{Corresponding author: School of Computer and Information Science, Southwest University, Chongqing 400715, China.  Tel:+86 023-68254555; E-mail address: ydeng@swu.edu.cn;prof.deng@hotmail.com}

\begin{abstract}
In many real complex networks, the fractal and self-similarity properties have been found. The fractal dimension is a useful method to describe fractal
property of complex networks. Fractal analysis is inadequate if only taking one fractal dimension to study complex networks. In this case, multifractal analysis of complex networks are concerned. However, multifractal dimension of weighted networks are less involved. In this paper, multifractal dimension
of weighted networks is proposed based on box-covering algorithm for fractal dimension of weighted networks (BCANw). The proposed method is applied to calculate the fractal dimensions of some real networks. Our numerical results indicate that the proposed method is efficient for analysis fractal property of weighted networks.
\end{abstract}


\end{frontmatter}


\section{Introduction}
Complex networks have been studied in various fields, including computer science, physics, management science,
biology, etc\cite{newman2003structure,amancio2011using,song2010synchronization,yang2011modified,vidal2011interactome,wei2013networks,zhang2013impulsive}.
Some fundamental properties of real complex networks have attracted much attention, such as small-world phenomena \cite{watts1998collective}, scale-free
degree \cite{barabasi1999emergence} and community structure \cite{fortunato2010community} etc. In 2005, fractal and self-similarity properties of complex networks have been investigated by Song et al \cite{song2005self}. The fractal dimension is used for fractal analysis of complex networks. The box-covering algorithm is described in detail and applied to calculate the fractal dimension of many real networks \cite{song2006origins,song2007calculate}. Subsequently, the classical box-covering algorithm for complex networks is extensively by many researchers \cite{shanker2007defining,gallos2007review,gao2008accuracy,schneider2012box,turnu2013fractal,hxzhang2013selfsimilarity}. However,
fractal analysis is inadequate when complex networks is studied by a single fractal dimension. In this case, multifractal analysis is a useful way to systematically
characterize the spatial heterogeneity of both theoretical and experimental fractal patterns \cite{dan2012multifractal,grassberger1983characterization,halsey1986fractal}. Multifractal has been applied successfully in
many different fields such as financial modeling \cite{anh2000multifractal,anh2000cointegration}, time series analysis \cite{canessa2000multifractality} and complex networks \cite{song2005self,long2009fractal,kim2007fractality}. For fractal analysis of complex networks, Wang et al.\cite{dan2012multifractal} introduced an improved box-covering algorithm for multifractal analysis of complex networks, a family of fractal networks is studied by Li et al. \cite{li2014fractal}. However, the existing works mainly focus on handling the fractal dimension of unweighted networks. An improved box-covering algorithm is used for fractal dimension of weighted networks (BCANw)in Ref \cite{wei2013box}. By adopting the weighted box-covering algorithm, the fractal property of weighted networks can be well described. In this paper, an algorithm of multifractal of weighted networks is proposed, and multifractal property of some real weighted networks is revealed.

In the following sections,  BCANw and multifractal method are introduced in section 2. The proposed model of multifractal analysis for complex networks is described in section 3. In section 4,  multifractal
dimensions of some real complex networks are calculated by the proposed method. Some conclusions are presented in section 5.

\section{Preliminaries}
\subsection{Box-covering algorithm for weighted networks}
In this section, a box-covering algorithm for weighted networks (BCANw) is briefly introduced. Given a weighted network $G={(N, V, W)}$, $N={(1,2,\cdots,n)}$ is a set of nodes and $V={(1,2,\cdots,m)}$ is a set of edges. $W={(1,2,\cdots,m)}$ is a set of edge-weight and denoted by $w_{ij}$.  $w_{ij}$ is value of edge-weight. The network $G$ is unweighted when the cell $x_{ij}(i,j=1,2,\cdots,n)$ of edge is
equal to 1 if node $i$ is connected to node $j$, and 0
otherwise. $G$ is defined as a weighted network if $w_{ij}$ could be any real
numbers. For unweighted networks, the shortest path between node $i$ and
node $j$ is defined as follows.
\begin{definition}[shortest path in unweighted networks]
Denoting $d_{ij}$ as the shortest path of between node $i$ and
node $j$, which satisfies
\begin{equation}\label{shortest}
d_{ij} = min{(x_{ih}+\cdots +x_{hj})}
\end{equation}
\end{definition}
There are two cases in weighted networks that needs notice. One case is that higher the weights, larger the distance of shortest path. The other is the opposite. Shortest path of the weighted networks is uniform defined as follows \cite{wei2013box}.
\begin{definition}[the shortest path in the weighted networks]
Denoting $d_{ij}$ as the shortest path of between node $i$ and
node $j$, which satisfies any of the following conditions
\begin{equation}\label{shortest1}
{d_{ij}} = \min (w_{i{j_1}}^p + w_{{j_1}{j_2}}^p +  \cdots +w_{{j_{m - 1}}{j_m}}^p + w_{{j_m}j}^p)
\end{equation}
where $j_{m}(m=1,2,\cdots)$ are IDs of nodes and $p$ is a real number.
\end{definition}
The weighted networks have different edge weights, which can be
non-integers. Thus, values of $d_{ij}$ may be non-integers too. It is also probable that the value of
$d_{ij}^{max}$ is less than one. values of the box size neither
increase by one in turn nor the initial value equal 1. Fractal property of weighted networks cannot be reversed by the classical box-covering algorithm of complex networks \cite{song2007calculate}. An improved box-covering algorithm for weighted networks is obtained in Ref \cite{wei2013box}. It is obtained by
stacking the value of distance until the value is more than the value of $d_{ij}^{max}(i,j=1,2,\cdots ,n)$.
The detailed algorithm of BCANw is given as follows.
Firstly, values of $d_{ij}$ between node $i$ and node $j$ connected directly are obtained. Suppose the order of $d_{ij}$ is denoted by  $d_{1} < d_{2} <\cdots< d_{m}$. And then, value of $l_{B}$ is a set and denoted by $D$, where $D={d_{1},d_{1}+d_{2},\cdots, \sum\limits_i^m {d_{i}}}$. Finally, for a given box size $l_{B}$, node $i$ and node $j$ are connected if ${d_{ij}} \ge {l_B}$. In the following steps, it is similar to the classical box-covering algorithm of complex network. The proposed box-covering algorithm for the weighted networks is a modified version of coloring algorithm. Value of box size is obtained by stacking the value of distance. It can directly calculate the fractal dimension of the weighted networks, and more corresponding points between box size and number are obtained.

\subsection{Multifractal method}
Multifractal analysis is a useful way to describe fractal property of complex system \cite{dan2012multifractal,grassberger1983characterization,halsey1986fractal,ivanov1999multifractality}.
For a set $E$ in a metric space, a probability measure $0 \le \mu  \le 1$ is given. A partition sum is considered as follows:
\begin{equation}
{Z_\varepsilon }(q) = \sum\limits_{\mu (b)} {{{[\mu (b)]}^q}}
\end{equation}
where $q$ is a real number and $\mu(b)$ is $\mu(.)$ of different non-overlapping boxes b which cover E with a give size $\varepsilon$. The mass exponent function of $\mu$ is denoted $\tau (q)$ and defined as follows:
\begin{equation}
\tau (q) = \mathop {\lim }\limits_{\varepsilon  \to 0} \frac{{\ln {Z_\varepsilon }(q)}}{{\ln \varepsilon }}
\end{equation}
The generalize fractal dimension of the measure $\mu$ is denoted $D_q$ which is defined as follows:
\begin{equation}
{D_q} = \frac{{\tau (q)}}{{q - 1}},q \ne 1
\end{equation}
Specially, ${D_1} = \mathop {\lim }\limits_{\varepsilon  \to 0} \frac{{{Z_{1,\varepsilon }}}}{{\ln \varepsilon }}$.
A view of life from the other side of multifractal, multifractal is defined as follows:
\begin{equation}
f(\alpha ) =  - \frac{{\ln N(\alpha )}}{{\ln \varepsilon }}
\end{equation}
where $N(\alpha)$ is a number of box in $[\alpha, \alpha+d\alpha]$. It means that $f(\alpha)$ is fractal dimension of all boxes with $\alpha$.
There is Legendre transform between $f(\alpha )$ and $\tau (q)$ \cite{mandelbrot1984fractal} as follows:
\begin{equation}
f(\alpha ) = q\alpha (q) - \tau (q)
\end{equation}
where $\alpha (q) = \frac{{d\tau (q)}}{{dq}}$

\section{Multifractal analysis of weighted networks}
\subsection{Multifractal algorithm of weighted networks}
Some algorithms of multifratal analysis are introduced \cite{yu2013fractal,li2014fractal,dan2012multifractal}.
For multifractal of complex networks, $\mu(b)$ is given as $\mu(b)=\frac{N_b}{N}$, where $N_b$ is the number of nodes covered by the box b and $N$ is the number of nodes. For a given box size $r$, the number of boxes is obtained by BCANw. And then, values of $\mu(b)$ is obtained. We calculate the partition sum as ${Z_r}(q) = {\sum\limits_{\mu (b) \ne 0} {[\mu (b)]} ^q}$ for each value of $r$.
In order to obtained the generalized fractal dimension $D_q$, lineae regression is an essential step for the appropriate range of box size. In this paper,
the linear regression of $ln {Z_r}(q)/(q - 1)$ against $ln(r/d)$ ($q \ne 1$) is considered. Last, values of $D_q$ are obtained for various values of $q$.
We determine the multifractality of weighted network by analyzing the shape of $D_q$.

\subsection{Multifractal analysis of some real weighted networks}
\begin{figure}[!t]
\centering
\subfigure[The USAir97 network] {\includegraphics[width=2.3in]{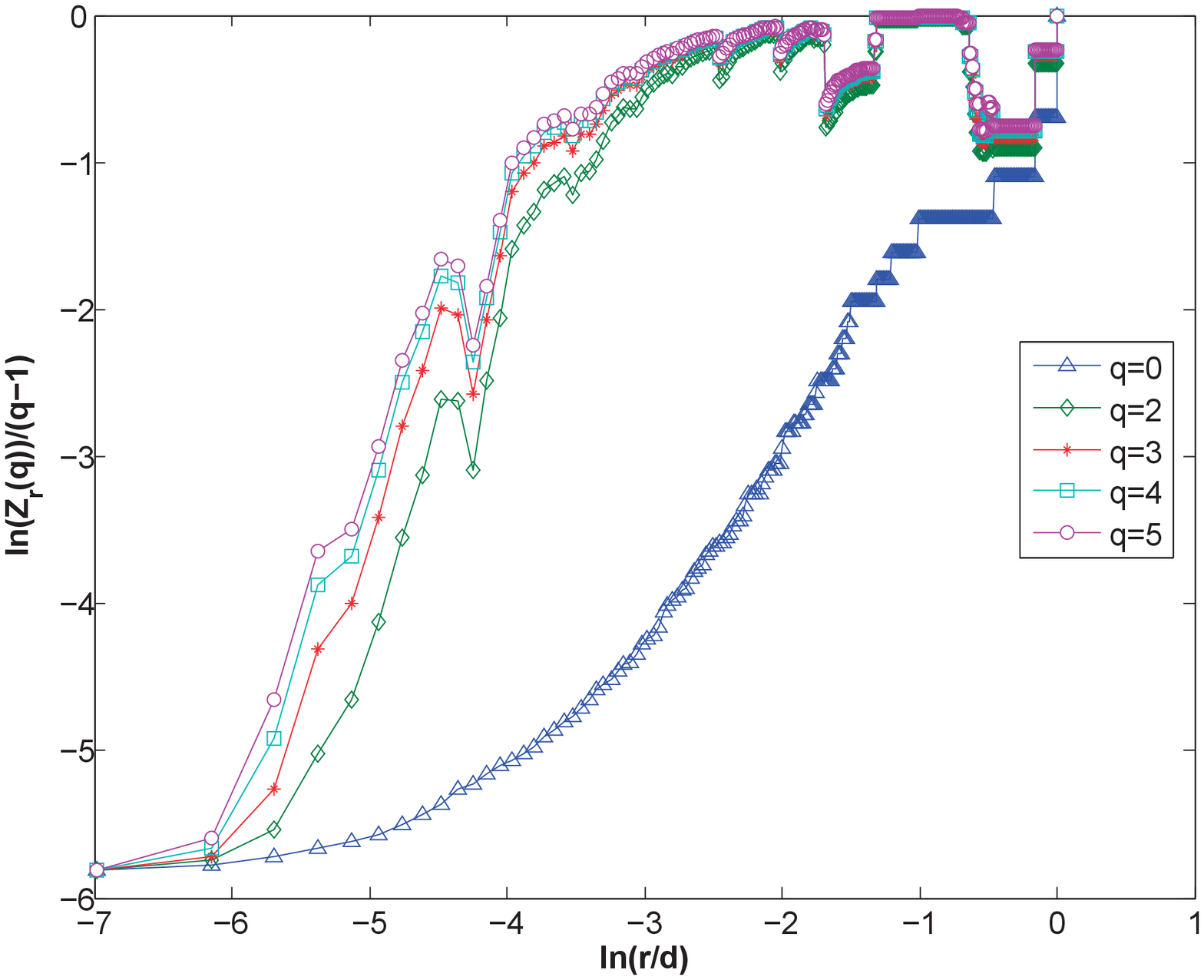}}
\subfigure[The C.elegans network] {\includegraphics[width=2.3in]{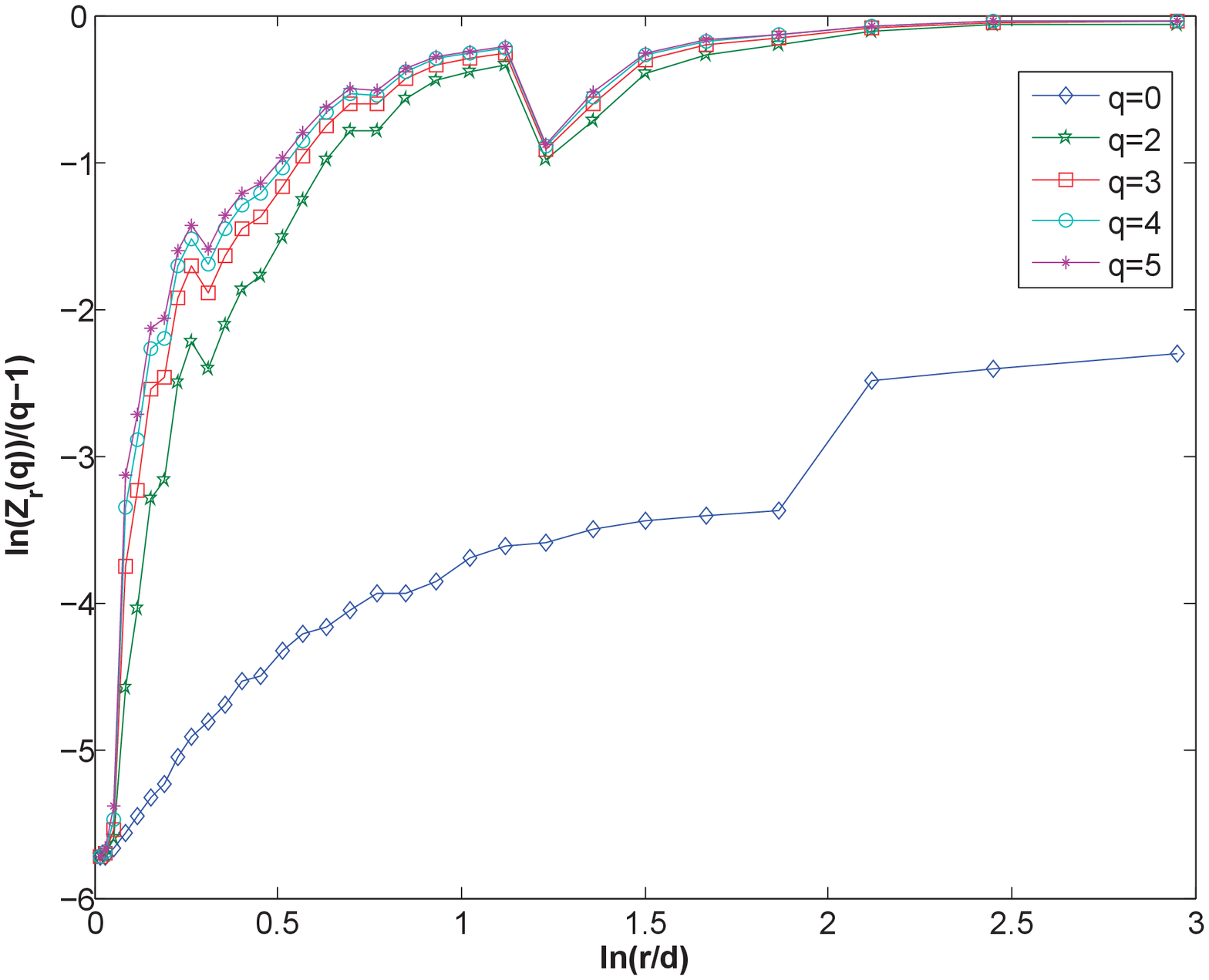}}
\subfigure[The Scientific collaboration network] {\includegraphics[width=2.3in]{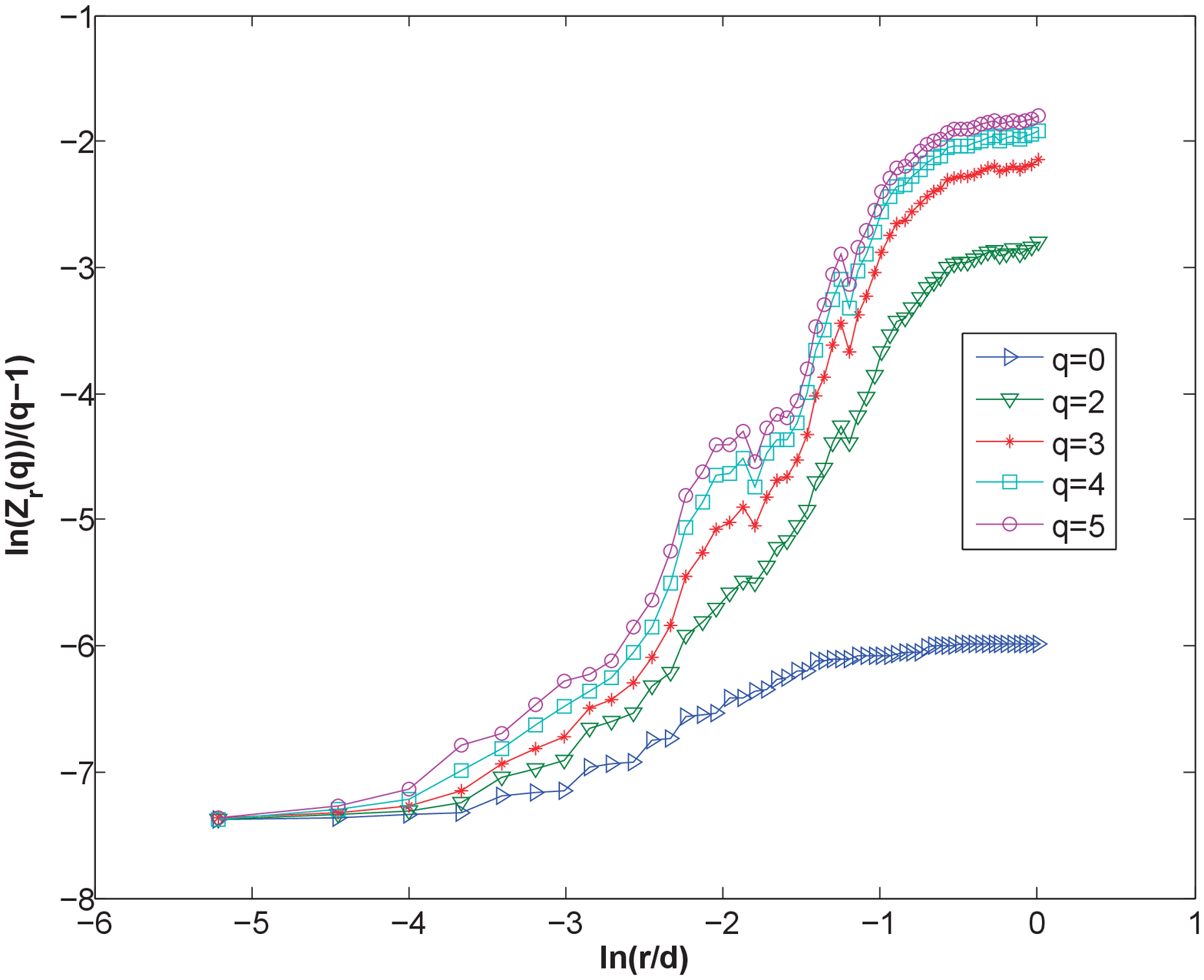}}
\caption{ Multifractal scaling analysis of these real complex networks. }
\label{fig}
\end{figure}

In the section, some real weighted networks - the USAir97 network, the C.elegans network and the Scientific collaboration network - are considered
The details of these weighted networks are obtained in Ref\cite{wei2013box}.  $ln {Z_r}(q)/(q - 1)$ against $ln(r/d)$ ($q \ne 1$) of these networks are considered, shown in Figure \ref{fig}.

\begin{figure}[!t]
\centering
\subfigure{\includegraphics[width=3.6in]{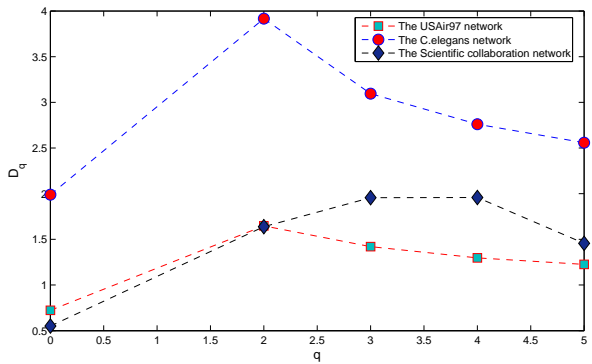}}
\caption{ The $D_q$ curves of the USAir97 network, the C.elegans network and the Scientific collaboration network. }
\label{fig1}
\end{figure}

In Figure \ref{fig}, the value range of $r$ is obtained by choosing appropriate points. Value of horizontal coordinates of USAir97 is [-6, -2.5]. In the C.elegans network and the Scientific collaboration network, we adopt the points from 11th to 30th and from 5th to 20th, respectively. According to these, the relationship between $D_q$ and $q$ is considered and shown in Figure \ref{fig1}.
In Figure \ref{fig1}, the values of $D_q$ in all the three weighted networks  increase when $q \in [0,2]$, and reach their peak values around $q=2$ for the USAir97 network and the C.elegans network. However, the maximum value of $D_q$ of the Scientific collaboration network is the interval $[3,4]$.

\section{Conclusions}
Recently, research on complex networks have shown that some real network exhibit the property of fractal scaling. Some different physical quantities are considered in the definitions of fractal including the multifractal analysis of complex networks. However, the existing stuies mainly deal with the fractal property of unweighted networks. In this paper, a multifractal analysis method for weighted networks has been proposed based on BCANw algorithm \cite{wei2013box}. The numerical example of real weighted networks shows that the proposed approach can well
reveal fractal property of weighted networks. To sum up, the proposed method is capable to reveal the fractal property of complex networks.

\section*{Acknowledgment}
The work is partially supported by National Natural Science Foundation of China (Grant Nos. 61174022, 71271061 and 61364030), National Key Technology R\&D Program (Grant No.2012BAH07B01), National High Technology Research and Development Program of China (863 Program) (No.2013AA013801), Chongqing Natural Science Foundation (for Distinguished Young Scholars) (Grant No. CSCT,2010BA2003), Doctor Funding of Southwest University (Grant No. SWU110021),.

\bibliographystyle{model1-num-names}
\bibliography{weighted}


\end{document}